# Multiple State Electrostatically Formed Nanowire Transistors

G. Segev, I. Amit, A. Godkin, A. Henning, Y. Rosenwaks

*Abstract*— Electrostatically Formed Nanowire (EFN) based transistors have been suggested in the past as gas sensing devices. These transistors are multiple gate transistors in which the source to drain conduction path is determined by the bias applied to the back gate, and two junction gates. If a specific bias is applied to the side gates, the conduction band electrons between them are confined to a well-defined area forming a narrow channel- the Electrostatically Formed Nanowire. Recent work has shown that by applying non-symmetric bias on the side gates, the lateral position of the EFN can be controlled. We propose a novel Multiple State EFN Transistor (MSET) that utilizes this degree of freedom for the implementation of complete multiplexer functionality in a single transistor like device. The multiplexer functionality allows a very simple implementation of binary and multiple valued logic functions.

## I. Introduction

ELECTROSTATICALLY Formed Nanowires (EFN) based transistors have been recently suggested as robust sensing[1] and memory devices [2]. The EFN device is based on the silicon-on-insulator (SOI) four gate field-effect transistor ($G^4$-FET) developed in 2002 [3], [4] which emerged from the volume inversion SOI MOSFET[5]. The $G^4$-FET combines MOSFET and junction gate field-effect transistor (JFET) principles as it consists of a top MOS gate ($V_{TG}$), a bottom substrate gate ($V_{BG}$) and is enclaved between by two lateral junction gates ($V_{JG1}$, $V_{JG2}$). The four gate transistor can be naturally adapted to CMOS technology scaling and manufactured in conventional silicon-on-insulator (SOI) processes with a low cost and high volume manufacturing. In EFN devices the top gate is removed in order to allow functionalization of the top surface, for example for gas sensing as suggested in [1], [6]. When a specific bias is applied to the side junction gates, the area around them becomes depleted and the conduction band electrons between the gates are confined to a well-defined area forming a narrow channel, the electrostatically-formed nanowire[7]. Glazman *et al.* proposed the lateral position control of an electron channel with a split-gate FET[8]. Shalev *et al.*[1] have shown in 3D electrostatic simulations that by applying non-symmetric bias on the side gates, the position of the EFN can be moved towards one of the gates.

We propose here novel Multiple-State EFN Transistors (MSET) that exploits the EFN lateral movement in order to form a single transistor multiplexer. In this device the drain is split into several isolated individual drains and the MSET output is defined by a narrow conduction channel between a specific drain and the source. The multiplexer functionality allows implementation of any logical operation within its inputs and outputs range. Furthermore, since it supports multiple well defined conduction states, the MSET can perform multiple valued logic (MVL) operations as well. Since this device is based on simple SOI concepts, it can be integrated into current technology relatively easily. To the best of our knowledge, this is the first CMOS compatible transistor for MVL large-scale circuits to be suggested. Although this work focuses on utilizing the MSET for logic applications, it is clear that it can be used in many other fields as well.

## II. Device operation

The basic MSET configuration closely resembles a split gate JFET with one important difference: the drain is split into several drain inputs, labeled $d_1..d_n$, isolated from each other by a dielectric buffer layer. A predefined combination of side gate voltages positions the EFN next to one of the drains while depleting the regions around all the other. As a result, a single conduction path between a single drain and the source is formed. For logic applications as discussed below, the bottom gate is redundant and can be removed. Fig. 1(a) shows an illustration of a two-drain MSET device embedded in a circuit which allows definition of voltage based states. The device bulk is n type doped and the source and drains are $n^+$ doped. The gates regions are $p^+$ doped. The light blue region between the drains is the dielectric buffer, e.g. $SiO_2$. When a voltage $V_1<0$ is applied to one of the gates and the other gate is grounded, the conduction channel will be next to the drain close to the grounded gate. In the example shown in Fig. 1, the applied biases are $V_{G,1}=V_1$ and $V_{G,2}=0$ such that the EFN is next to $d_2$. Since the resistance of the conduction path between $d_2$ and the source is significantly lower than the resistance of the serial resistor $R$, the output voltage is $V_s=V_{d,2}$. In a similar fashion, applying opposite voltages, $V_{G,1}=0$ and $V_{G,2}=V_1$ yields $V_s=V_{d,1}$. When both gates are at voltage $V_2<0$ the entire region between the two gates is completely depleted. As a result, the resistance between the source and the two drains is higher than R and $V_{out}$ is 0. When both gates are at 0 there is conduction between both drains and the state is undefined. This intermediate state can increase the device functionality considerably but may result in high currents between the drains and high power consumption. It should be noted that similar to the CMOS configurations, in the proposed configuration only leakage current flows through the device once the state is determined. In order to further reduce the source current, the resistor in Fig.1 can be replaced with transistors that will conduct or isolate the path between $V_{out}$ and the ground according to the desired functionality.

Adding a third drain to the MSET greatly increases its functionality. With three possible conduction states, this device can perform ternary based logical operations. In this configuration, a high negative voltage on one gate and zero on the other will allow conductance from one of the lateral drains and a moderate negative voltage on both side gates allows conductance from the middle drain.

In order to obtain additional degrees of freedom from the MSET a three dimensional architecture is proposed. The basic mode of operation is a two dimensional movement of the channel, where the conduction path is determined by 4 independent gates, termed here $JG_N$, $JG_S$, $JG_E$ and $JG_W$. The drains are `islands', isolated from each other and located between the gates. The source is located beneath the drains. An illustration of a three dimensional MSET with four gates and four drains is shown in Fig 1(b). In this case the conductence column is determined by the voltage on $JG_W$ and $JG_E$, and the conducted row is determined by the voltages on $JG_N$ and $JG_S$.

The advantages of the MSET can be easily demonstrated by realizing a 2 control, 4 inputs multiplexer as illustrated in Fig. 1(b). Here the inputs $a_0$ and $a_1$ are binary signals that determine the voltage on $JG_W$ and $JG_N$. Two CMOS inverters are used to transmit the complementary signal to the two opposite gates. As a result, for each combination of $a_0$ and $a_1$ there will be a single conduction path between one of the drains and the output. It should be noted that while this configuration consists of a single MSET and 4 MOSFETs, a similar none-restoring multiplexer realized in CMOS consists of 16 transistors [9].

In order to realize complex logical functions, several devices must be connected in series. To this end, the MSET output voltage (source voltage) must be of the same sign and magnitude as the input voltage (gates and drain voltage).

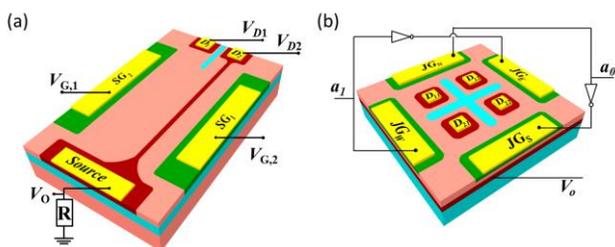

Fig. 1. (a) two drains MSET device in device in an exemplary circuit which allows definition of voltage based states. An exemplary realization of a 2 control 4 inputs multiplexer based on a four gates, four drains MSET.

However, in order to avoid forward biasing between the gates-drains p-n junction the drain voltages must be non negative and the gate voltages must be non positive. This contradiction can be resolved in several ways: addition of a voltage shifting circuit, addition of a voltage inverting circuit or alternating p-n design where the output of an n type MSET is connected to the gates of a p type MSET. Obviously, such a design requires that the drain voltages of the n type MSET will be exactly the required gate input of p type MSETs and vies-versa.

### III. SIMULATION

In order to demonstrate the basic MSET concept of operation, a proof-of-concept two-drains MSET was simulated with Sentaurus TCAD. The circuit is as in Fig 1(a) where the resistor is of 10MΩ and the drain voltages $V_{d,1}$ and $V_{d,2}$ are 0.75V and 0.5V respectively. The MSET bulk doping is $10^{17}cm^{-3}$ n type, the gate regions are $5 \cdot 10^{19}cm^{-3}$ p type doped and the source and drains are $5 \cdot 10^{19}cm^{-3}$ n type doped. The geometry of the simulated device is shown in Fig 2(a), all the geometrical parameters are scaled evenly. A contour of the source voltage as a function of the two gate voltages is shown in Fig 2(b). The voltage difference between any two consecutive iso-voltage lines is 25mV. The dashed lines show the regions where the output is within 1% of the appropriate drain voltage. When a high voltage is applied to one of the side gates the conduction path is pushed away from it forming a single channel between the opposite drain and the source. As a result, when $V_{g,2}$ is more negative than -0.75V and $V_{g,1}$ is close to 0V, the output voltage is nearly $V_{d,1}$. Similarly, when $V_{g,1}$ is below -0.75 and $V_{g,2}$ is close to zero, the output voltage is close to $V_{d,2}$. Last, when both gates are lower than -1.5V, there is no conduction through the device and the output is close to 0V. Hence, as can be seen by the areas surrounded by the dashed curve, three different states can be defined for this device:, $S_1$ where $V_o=V_{d,1}$ $S_2$ where $V_o=V_{d,2}$ and $S_3$- where $V_o=0$.

As discussed above increasing the number of drains can increase the device functionality significantly. The next device simulated is a 2 gates, 3 drains MSET. The circuit is as in Fig 1(a) where the resistor is of 1GΩ and the drain voltages $V_{d,1}$, $V_{d,2}$ and $V_{d,3}$ are 0.75V, 0.5V and 0.25V respectively. The geometry of the simulated device is shown in Fig 2(c), all the geometrical parameters are scaled evenly. The device doping concentrations are as in the 2 drains MSET apart for the N$^-$ region which is n type doped with a concentration of $10^{16}$ cm$^{-3}$. A contour of the source voltage as a function of the two side gates voltages is shown in Fig. 2(d). The voltage difference between two iso-voltage lines is 25mV. The dashed lines show the regions where the output is within 1% of the appropriate drain voltage. As in the previous simulation, the different states can be easily distinguished. However, in this configuration the displacement of the conduction channel must be larger in order to allow conductance from each single drain separately. As a result, the required voltages are higher than in the 2 drains simulation. Furthermore, depletion between the gates and the drains results in voltage drops between the drains and source limiting the maximum drain voltages.

Hence, further optimization is required in order to have matching gates and drains voltages which allow device concatenation.

## IV. CONCLUSION

The proposed class of devices is a proof of concept aimed to demonstrate their basic functionality. Wide spread implementation of MSET based circuits in ASIC and FPGA chips will rely on the ability to fabricate them in appropriate lateral dimensions with low power consumption at sufficient frequency. As discussed above, an MSET based 4 inputs multiplexer can be realized with a single MSET and 4 MOSFET transistors, comparing to 16 MOSFETs in CMOS technology. Hence, even if a 4 gates, 4 drains MSET is 12 times less efficient than current transistors in terms of size, power consumption and speed, the suggested MSET based circuit may still be more attractive. A second issue that has to be addressed is device concatenation. As discussed above, in order to concatenate MSET outputs with MSET gate inputs the drains and gates voltages must be of similar magnitudes allowing use of voltage shifting or inverting circuits. This demand must reconcile with the fact that the conduction channel movement range defines the required gate voltages in these devices. However, the drains voltages are limited due depletion between the drains and the nearby gates similar to the pinch-off effect in standard JFET transistors [10]. Reducing the magnitude of the drains and buffers along with detailed optimization of the device geometry and doping profiles can reduce these apparent contradictions considerably.

In order to fully exploit the functionality of MSET devices the number of states (drains) should be maximized. This number is limited by the fabrication process and the range of motion of the conduction channel for a specific range of gates voltages. Since the number of possible inputs of a 4 Gates MSET is the square of the number of drains, there is great motivation to increase the amount of drains to larger numbers than the those that were presented in this work. Although we have discussed the use of MSETs for implementation of logic operations, this family of transistors may prove beneficial in many analog and digital applications where simple multiplexing circuits are required.

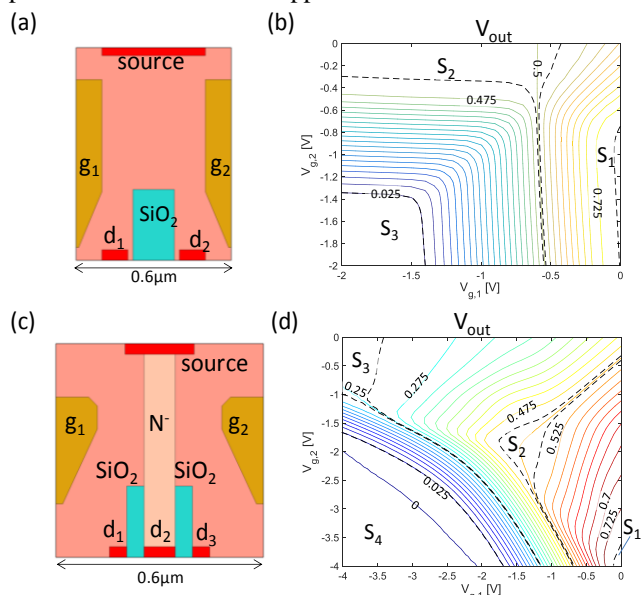

Fig. 2. The simulated MSETs geometry two drains MSET (a) and 3 drains MSET (c). The bulk doping is $10^{17}$cm$^{-3}$ n type, the gate regions are $5 \cdot 10^{19}$cm$^{-3}$ p type doped and the source and drains are $5 \cdot 10^{19}$cm$^{-3}$ n type doped. The N$^-$ region In the 3 drains MSET is $10^{16}$cm$^{-3}$ doped. All the geometrical parameters are scaled evenly. (b) The 2 drains MSET(b) and 3 drains (d) circuit output as a function of the gates voltages. The resistor between the source and the ground is of 10MΩ and 1GΩ in the two and three drains MSET circuit respectively. The drains voltages are 0.75 and 0.5V respectively in the 2 drains MSET and 0.75, 0.5V and 0.25V respectively in the three drains MSET. The voltage drop between two iso-voltage lines is 25mV. The dashed lines show the regions where the output is within 1% of the appropriate drain voltage